# A Better Definition of the Kilogram

by
Ronald F. Fox and Theodore P. Hill

9/16/07

Abstract:

Fixing the value of Avogadro's constant, the number of atoms in 12 grams of carbon-12, at exactly $84446886^3$ would imply that *one gram is the mass of exactly $18 \times 14074481^3$ carbon-12 atoms*. This new definition of the gram, and thereby also the kilogram, is precise, elegant and unchanging in time, unlike the current 118-year-old artifact kilogram in Paris and the proposed experimental definitions of the kilogram using man-made silicon spheres or the watt balance apparatus.

Article:

Since 1889, the official scientific definition of the basic unit of mass has been that one kilogram is the mass of *Le Gran K*, a unique platinum-iridium cylinder stored in a vault near Paris. Due to cleaning, handling and calibrating with other kilogram standards, however, the mass of that cylinder artifact is known to be changing in time [1, 4]. Unfortunately, this implies that the official mass of a single atom of carbon-12 (and every other atom in the universe) is changing in time. As Physicist Richard Davis of the International Bureau of Weights and Measures said [1], "We could obviously use a better definition" of the kilogram.

Proposals to redefine the kilogram experimentally using manmade silicon spheres and the watt balance apparatus [6,7] suffer from the same problem as Le Gran K – the experiments are inherently inexact, and the results are also changing in time depending on the equipment and laboratory.

A better definition of the kilogram has been described [2, 3] using the relationship between the kilogram and the definition of the mole via Avogadro's number. Fixing the value of Avogadro's constant, the number of atoms in 12 grams of carbon-12, at exactly $84446886^3$ would imply that

   *one gram is the mass of exactly $18 \times 14074481^3$ carbon-12 atoms.*

Simply fixing the value of a kilogram numerically is analogous to past decisions by the scientific community to fix, once and for all, the value of a *second* at exactly 9,192,631,770 vibrations of a particular hyperfine transition between two states of a cesium-133 atom, and the *meter* as the distance light travels in exactly 1/299,792,458 seconds. That decision, on October 21, 1983, eliminated the need for the official artifact platinum-iridium meter stick forever [2]. Fixing an exact value for Avogadro's number has been endorsed by the Committee on Nomenclature, Terminology and Symbols of the American Chemical Society [5], and the particular value chosen, $84446886^3$, is consistent with current accepted values for all other fundamental physical constants and theories. The argument for fixing this number as a perfect cube is given in [2].

This new definition of the kilogram is precise, elegant and unchanging in time, unlike the 118-year-old artifact kilogram in Paris and the proposed experimental definitions of the kilogram. It would not only have the advantage of immediately replacing the current kilogram artifact, but also would have the advantage of making the definitions of the atomic mass unit and the mole explicit, clean and simple:

   1 *amu* is exactly $1/(18\times 14074481^3)$ gram; and

   1 *mole* of any entity (element, chemical compound, etc) is exactly $84446886^3$ of those entities.

In the words of the Chair of the Committee on Nomenclature, Terminology and Symbols of the American Chemical Society, Professor Paul Karol, by simply fixing an integral value for Avogadro's number in this manner, "much of what seems to confuse many students about the mole in introductory courses will be dampened." [5].

*Le Système International d'Unites* (SI), the organization that oversees measurements and standards that have been officially recognized and adopted by nearly all countries, identifies exactly seven basic units and their standards: length (meter), mass (kilogram), time (second), electrical current (ampere), thermodynamic temperature (kelvin), amount of substance (mole) and luminous intensity (candela). Of these basic seven, which are assumed to be mutually independent, the kilogram is the only unit that is still defined in terms of a physical artifact, and the National Institute of Standards and Technology agrees [6] that is indeed time to find a better definition for the kilogram. The clean, simple, permanent mathematical definition above is an excellent candidate.

Ron.fox@physics.gatech.edu
hill@math.gatech.edu